\documentclass[12pt]{iopart}

\begin{document}

\title[Comments on the article "M.O.\,Katanaev, Complete separation of variables in ...'']{Comments on the article "M.O.\,Katanaev,\\ 
Complete separation of variables in the geodesic Hamilton--Jacobi equation in four dimensions, 
Physica~Scripta (2023), 98, 104001"}

\author{V~V~Obukhov$^{1,4}$, 
K~E~Osetrin$^{2,4}$,
and A~V~Shapovalov$^{3,4}$}

\address{$^1$ Institute of research and development, Tomsk State Pedagogical University,
Kievskaya str.~60, 
Tomsk~634061, Russia}

\address{$^2$ Center for Mathematical and Computer Physics, Tomsk State Pedagogical University,
Kievskaya str.~60, 
Tomsk~634061, Russia}

\address{$^3$ Department of Theoretical Physics, 
National Research Tomsk State University, Lenina pr.~36, Tomsk~634050, Russia}

\address{$^4$ International laboratory of theoretical cosmology, 
Tomsk State University of Control Systems and Radioelectronics, Lenina pr.~40, Tomsk~634050, Russia}


\ead{obukhov@tspu.edu.ru}

\begin{abstract}
The present note is a feedback on the article by M.O.\,Katanaev in  {\it Physica~Scripta (2023, 98, 104001)}, where, in our opinion, a distorted view of the classical theory of separation of variables in the Hamilton-Jacobi
equation is given. 
We show that the metrics given in this paper, addmiting separation of variables, are special cases of 
V.N.\,Shapovalov metrics {\it (Siberian Mathematical Journal, 1979, 20, 790)}, finally obtained in the 70s of the last century.
The results of the article in question, unlike the original ones, do not have scientific novelty and contain omissions and incorrect statements.
\end{abstract}

%
\vspace{2pc}

\submitto{\PS}

\maketitle

In a recently published article \cite{1}, as well as in the preprint \cite{2}, which is cited in this article, the author sets out his vision of the history of the creation of the theory of complete separation of variables in the free Hamilton-Jacobi equation and assesses the completeness of this theory. He claims that the results obtained in the articles significantly complement the theory. Thus, in the work \cite{2} the discovery of a new class of separable spaces (according to the currently widely accepted terminology - {St{\"{a}}ckel} spaces) was announced, which have zero diagonal components of the contravariant metric tensor, corresponding to isotropic non-ignorable variables of the separable coordinate system (that is, the coordinate system, in which separation of variables is possible). It is also stated that for the first time an explicit form of integrals of motion and complete integrals for the free Hamilton-Jacobi equation has been found. According to the author, it was in the articles \cite{1,2}, that the complete list of nonequivalent metrics of four-dimensional {St{\"{a}}ckel} spaces was first presented and also their new classification is proposed.

In the present comment we show that the papers \cite{1,2}, do not obtain any new results and do not contain any new ideas, since the papers themselves do not go beyond the definitions and consequences of the main theorems of the theory of complete separation of variables. These theorems have been proved half a century ago by Vladimir N. Shapovalov. This also applies to the complete list of metrics, integrals of motion and complete integrals of the free Hamilton-Jacobi equation in separable coordinate systems for an arbitrary $4$-dimensional {St{\"{a}}ckel} space $V_4$ given in the article \cite{1}. Such a list (though a little more complete) has also been known to specialists for a long time

The problem of separation of variables in the Hamilton-Jacobi equation for geodesics is more than a hundred years old. It was first formulated in the papers of Paul {St{\"{a}}ckel} in the late 19th century. Paul {St{\"{a}}ckel}, using the definition of complete separation of variables according, proposed a method of constructing metrics of spaces in which the geodesic equations can be integrated by the method of complete separation of variables in the free Hamilton-Jacobi equation, and also found the first examples of metrics of spaces, in which such separation of variables takes place. In the separable coordinate system these metrics are diagonal, and the Hamilton-Jacobi equation admits quadratic on momentum integrals of motion.
Subsequently, many articles were devoted to the problem of complete separation of variables. A detailed bibliography can be found in the papers \cite{3,4,5,6}. Until the early seventies of the 20th century, non-isotropic metrics of spaces were found in which the complete separation of variables in the free Hamilton-Jacobi equation can be used, and it was also established that the integrals of motion are constructed using the {St{\"{a}}ckel} method through the {St{\"{a}}ckel} matrix introduced by him. The method is applicable to non-diagonal metrics, provided that the Hamilton-Jacobi equation, along with integrals of motion quadratic in momenta, admits a commutative algebra of integrals of motion linear in impulse. In this case, the space metric is determined, in addition to the elements of the {St{\"{a}}ckel} matrix, by an additional set of functions, each of which depends only on one non-ignored variable of the separable coordinate system. It was proposed to call all spaces in which the modified {St{\"{a}}ckel} method is applicable {St{\"{a}}ckel} spaces.
It was also found that a necessary condition for the complete separation of variables in the free Hamilton-Jacobi equation is the presence in the n-dimensional Riemannian space $V_n$ with an arbitrary signature of a set of $n$ geometric objects, consisting of mutually commuting vector and tensor Killing fields, including the metric tensor (equivalently - a complete set of linear and quadratic in momentum integrals of motion of the second kind for the free Hamilton-Jacobi equation). Note that all these results were obtained in non-covariant form in the separable coordinate system. Isotropic {St{\"{a}}ckel} metrics in the separable coordinate systems were first proposed in \cite{7,8,9}.

The final construction of the theory of complete separation of variables in the free Hamilton-Jacobi equation was carried out in covariant form by Vladimir N. Shapovalov  (see \cite{9} and other earlier articles). In these articles, the main theorems of the theory of {St{\"{a}}ckel} spaces were proved. The theorems establish necessary and sufficient conditions for the complete separation of variables in the free Hamilton-Jacobi equation. One of these theorems explicitly provides comprehensive formulas for the general canonical form  of {St{\"{a}}ckel} spaces  metrics and of integrals of motion for free Hamilton-Jacobi equation in the separable coordinate system. Thus, the problem of listing all metrics of spaces in which the method of complete separation of variables is applicable is completely solved. Moreover, according to Shapovalov's fundamental theorem, the equivalence classes of {St{\"{a}}ckel} spaces are uniquely defined by complete sets of Killing fields. 
Therefore, there is a tool that allows one to determine in any coordinate system whether the space under consideration belongs to the class of {St{\"{a}}ckel} spaces. Papers \cite{1,2}, do not go beyond the theory of complete separation of variables built fifty years ago and does not contain any new results. In order to verify this, we present the canonical form of metrics of {St{\"{a}}ckel} spaces in separable coordinate systems proposed by Vladimir N. Shapovalov in his work \cite{9}. In this paper, the problem of partial separation of variables was studied in more general form. 
The problem of complete separation of variables is also considered and completely solved as a special case of this more general problem. Therefore, in order to simplify the formulas, we will slightly change the notation adopted in \cite{9}.
In what follows, the next indices of the variables of the separable coordinate system in the $n$-dimensional {St{\"{a}}ckel} space $V_n$ will be used.
$$
i,j = 1, .. , n; \quad p, q, r = 1, .. , N; \quad \nu,\mu = N+1, .. , n;
$$

$$
\nu_0,\mu_0 = N+1, .. , N+N_0,\quad \nu_1,\mu_1 = N+N_0 +1, ... , n; \quad 0 \leq N_0 \leq N \leq n.
$$
Let us consider the Hamilton-Jacobi equation in the separable coordinate system of the $n$-dimensional plug space $V_n$.
\begin{equation}\label{1}
H=g^{ij}p_ip_j = \lambda_n, \quad  p_i=\partial S/\partial u^i
\end{equation}
The equation (\ref{1}) admits a complete set of mutually commuting integrals of motion:
\begin{equation}\label{2}
 X_q = X_p^i p_i, \quad  X_\nu=X^{ij}_\nu p_ip_j, \quad [X_i,X_j] = 0  \quad (H=X_n).
\end{equation}
According to Shapovalov theorem, the components of tensors $X^{ij}_\nu$ in the separable coordinate system can be represented as
\begin{equation}\label{3}
g^{ij} = h^{ij}_\nu (\Phi^{-1})^\nu_n, \quad X^{ij}_\mu = h^{ij}_\nu (\Phi^{-1})^\nu_\mu \quad n = dim V_n.
\end{equation}
where the functions $h_\nu^{ij}(u^\nu), \quad \Phi^\mu_\nu(u^\nu)$ depend only on the variables
\begin{equation}\label{4}
h^{ij}_\nu =\delta^i_{\nu_1} \delta^j_{\nu_1} + (\delta^i_{\nu_0}\delta^j_p  + \delta^i_p\delta^j_{\nu_0})h^{p\nu{_0}} + \delta^i_p\delta^j_p h^{pq}_{\nu}
\end{equation}
In the separable coordinate system, the full integral of the equation (\ref{1}) can be represented as:
\begin{equation}\label{5}
S = \Sigma\varphi_i(u^i, \lambda_i), \quad \lambda_i = const.
\end{equation}
Functions $\varphi_i$ are defined in quadrature from the system of equations:
$$
X_p= X^i_p p_i = \delta_p^i p_i =\lambda_p\Rightarrow \quad \varphi_{p,p} = \lambda_p \quad \Rightarrow \quad \varphi_p = \lambda_p u^p, \quad
$$
$$
(\varphi_{\nu_1,\nu_1})^2 = \Phi^\mu_{\nu_1}\lambda_\mu - h^{pq}_{\nu_1}\lambda_p\lambda_q, \quad\varphi_{\nu_0,\nu_0} = \frac{\Phi^\mu_{\nu_1}\lambda_\mu - h^{pq}_{\nu_0}\lambda_p\lambda_q}{2h^{r{\nu_0}}_{\nu_0}\lambda_r},
$$
which are integrated in quadrature. Note that the number $N_0$ is invariant since it is calculated by the formula:
$$
N_0 = N-rank||X_p^ig_{ij} X_q^i||
$$
Thus the type of the complete separation of variables and the type of the {St{\"{a}}ckel} space are completely given by the dimension of the space, and the numbers $N$,  $N_0$ characterizing the abelian algebra included in the complete set. According to the terminology adopted for four-dimensional {St{\"{a}}ckel} spaces  they are called {St{\"{a}}ckel}  spaces of type $(N, N_0)$. If $N_0 =0$, the spaces are called non-isotropic (spaces of non null types). Otherwise, them are called isotropic (spaces of null types). As it is easy to calculate, there are only 9 non-equivalent types of {St{\"{a}}ckel}  metrics in $V_4$. In the paper  \cite{1}  a list of 10 metrics is provided. However, two of them (formulas (12), (13) in the paper \cite{1}) should be combined since they are equivalent. Both of them are special cases of canonical metrics of type $(2.1)$ according to the classification of Shapovalov (see \cite{4}). The same is true for the metrics (17), (18) in the paper \cite{1}. They are equivalent to each other and both are special cases of metrics of type $(1.1)$ according to Shapovalov classification (see \cite{5}). Note that the list of metrics completely lacks metrics of isotropic {St{\"{a}}ckel} spaces of type $(3.1)$. The canonical form of the metric tensor $g^{ij}$ for metrics of type $(3.1)$ in the separable coordinate system has the form.
\begin{equation}\label{6}
g^{ij} = h^{pq}(u^4)\delta_p^i\delta_q^j +(\delta_3^i\delta_4^j + \delta_4^j\delta_3^j) \quad (p,q = 1,2,3).
\end{equation}

After the final construction of the theory of complete separation of variables in the free Hamilton-Jacobi equation, a large number of papers (see, for example, \cite{3,4,5,6} and the literature cited in them) investigated the possibilities of applying the metrics (\ref{2}) in the special theory of relativity and in the theory of gravitation.  In particular, when solving the problem of separation of variables in the classical and quantum equations of motion in the presence of external fields (electromagnetic fields, Yang-Mills fields). Note also that long before the publication of \cite{1,2}, four-dimensional isotropic {St{\"{a}}ckel} metrics of all types were studied in detail in many papers (see, for example, \cite{4,5}, \cite{10,11,12}).

\section*{Conclusion}


Summarising, we believe that the statement made in the papers \cite{1, 2}
about new scientific results in the problem of complete separation of
variables in the Hamilton-Jacobi equation for geodesics does not correspond to
reality. We did not find any new results in comparison with those contained in
the theorem on necessary and sufficient conditions of Vladimir N. Shapovalov.
The results of the article in question, unlike the original ones, have
omissions, and the article contains incorrect statements. Thus, the methods
and results proposed in the works \cite{1, 2} do not contain scientific
novelty and represent a special case of the theory of complete separation of
variables in the free Hamilton-Jacobi equation, finally constructed in the 70s
of the last century. 

\section*{Statements and Declarations}

\subsection*{Data availability statement}

All necessary data and references to external sources are contained in the text of the manuscript. 
All information sources used in the work are publicly available and refer to open publications in scientific journals and textbooks.

\subsection*{Compliance with Ethical Standards}

The authors declare no conflict of interest.

\section*{Bibliography}


\providecommand{\newblock}{}

\end{document}